\documentstyle[12pt]{article}
\catcode`\@=11

\@addtoreset{equation}{section}
\def\theequation{\arabic{section}.\arabic{equation}}

\def\@normalsize{\@setsize\normalsize{15pt}\xiipt\@xiipt
\abovedisplayskip 14pt plus3pt minus3pt%
\belowdisplayskip \abovedisplayskip
\abovedisplayshortskip  \z@ plus3pt%
\belowdisplayshortskip  7pt plus3.5pt minus0pt}

\def\small{\@setsize\small{13.6pt}\xipt\@xipt
\abovedisplayskip 13pt plus3pt minus3pt%
\belowdisplayskip \abovedisplayskip
\abovedisplayshortskip  \z@ plus3pt%
\belowdisplayshortskip  7pt plus3.5pt minus0pt
\def\@listi{\parsep 4.5pt plus 2pt minus 1pt
            \itemsep \parsep
            \topsep 9pt plus 3pt minus 3pt}}

\def\underline#1{\relax\ifmmode\@@underline#1\else
        $\@@underline{\hbox{#1}}$\relax\fi}
\@twosidetrue

\relax

\catcode`@=12

\catcode`\@=11

\def\section{\@startsection{section}{1}{\z@}{3.5ex plus 1ex minus
   .2ex}{2.3ex plus .2ex}{\large\bf}}

\def\thesection{\Roman{section}.}

\def\appendix{\setcounter{section}{0}
        \def\thesection{APPENDIX }
        \def\theequation{\Alph{section}.\arabic{equation}}}
\def\ps@headings{\def\@oddfoot{}\def\@evenfoot{}
\def\@evenhead{\@oddhead}
\def\subsectionmark##1{\markboth{##1}{}}
}

\catcode`\@=12

\relax

\def\figcap{\section*{Figure Captions\markboth
        {FIGURECAPTIONS}{FIGURECAPTIONS}}\list
        {Fig. \arabic{enumi}:\hfill}{\settowidth\labelwidth{Fig. 999:}
        \leftmargin\labelwidth
        \advance\leftmargin\labelsep\usecounter{enumi}}}
 \relax
\def\tablecap{\section*{Table Captions\markboth
        {TABLECAPTIONS}{TABLECAPTIONS}}\list
        {Table \arabic{enumi}:\hfill}{\settowidth\labelwidth{Table 999:}
        \leftmargin\labelwidth
        \advance\leftmargin\labelsep\usecounter{enumi}}}
 \relax
\def\reflist{\section*{References\markboth
        {REFLIST}{REFLIST}}\list
        {[\arabic{enumi}]\hfill}{\settowidth\labelwidth{[999]}
        \leftmargin\labelwidth
        \advance\leftmargin\labelsep\usecounter{enumi}}}
 \relax

\catcode`\@=11

\def\ps@headings{\def\@oddfoot{}\def\@evenfoot{}
\def\@evenhead{\@oddhead}
\def\subsectionmark##1{\markboth{##1}{}}
}

\ps@headings

\relax
\newskip\humongous \humongous=0pt plus 1000pt minus 1000pt
\def\caja{\mathsurround=0pt}
\def\eqalign#1{\,\vcenter{\openup1\jot \caja
        \ialign{\strut \hfil$\displaystyle{##}$&$
        \displaystyle{{}##}$\hfil\crcr#1\crcr}}\,}
\newif\ifdtup

\def\beq{\begin{equation}}
\def\eeq{\end{equation}}

\def\beqn{\begin{eqnarray}}
\def\eeqn{\end{eqnarray}}
\relax

\def\G2{{\; \rm GeV/}c^2}
\def\G{\; \rm GeV}

\def\dotx{\dotx{\dot\overline{x}}}

\relax

\begin{document}
\hbadness=10000
\begin{titlepage}
\nopagebreak
\begin{flushright}

        {\normalsize KUCP-38\\
         PURD-TH-91-09\\

        September,~1991}\\
\end{flushright}
\vfill
\begin{center}
{\large \bf S-matrix of
N=2 Supersymmetric Sine-Gordon Theory}
\vfill
\renewcommand{\thefootnote}{\fnsymbol{footnote}}
{\bf Ken-ichiro Kobayashi}\footnote{Now at Institute for Nuclear Study,
University of Tokyo, Tanashi, Tokyo 188, Japan}

       Department of Physics, Purdue University, \\
       West Lafayette, IN 47907, USA \\
\vfill
       {\bf Tsuneo Uematsu}

       Department of Physics, College of Liberal Arts and Sciences, \\
       Kyoto University,~Kyoto 606,~Japan \\
\vfill

\end{center}

\vfill
\nopagebreak
\begin{abstract}
We investigate the S-matrix of N=2 supersymmetric sine-Gordon theory
based on the N=2 supersymmetry and the quantum group structure.  The
topological charges play an important role to derive physical contents.
\end{abstract}
\vfill
\end{titlepage}
\pagestyle{plain}
\newpage
\voffset = -2.5 cm

  Recently there has been much interest in 2-dimensional massive integrable
field theories that can be treated as perturbed conformal field theories [1].
Sine-Gordon theory is such a class of models that exhibits factorizability of
S-matrices.  It has been known that conformal minimal models perturbed by the
$\Phi_{(1,3)}$ operartor lead to the restricted sine-Gordon theory at rational
values of the coupling constant [2-6]. There, quantum group symmetry [7,8]
plays
an important role for truncating the Hilbert space as well as for deriving
S-matrices [2-5,9].

  Now it is interesting to see whether incorporation of supersymmetry will
affect the factorizable structure of S-matrices in sine-Gordon theory.
The N=1 sine-Gordon theory has been studied as the minimal superconformal
models perturbed by the $\Phi_{(1,3)}$ operator [10,11]. Recently N=2
supersymmetric sine-Gordon theory has been proposed and its conservation laws
were investigated at the
quantum level [12,13] as well as at the classical level [14].  It has turned
out [12,13] that the theory has commuting supersymmetry and
$\widehat{sl_{q}(2)}$ quantum group symmetry [15].
Based on the quantum structure, in this paper we propose the S-matrix of the
theory.  In the supersymmetric models studied so far, including N=1
supersymmetric sine-Gordon theory, it was realized that the factorization of
S-matrix into the bosonic part and supersymmetric factor is characteristic
[16,17,11].
In our N=2 case the factorization also holds from the two commuting
algebras.

  The classical equations of motion of N=2 sine-Gordon theory are
$$\eqalign{
& {\overline D}_{+}D_{+}\phi^{+} = g \sin \beta\phi^{-}  \cr
& {\overline D}_{-}D_{-}\phi^{-} = g \sin \beta\phi^{+}  \cr
}\eqno(1)$$
where the N=2 supercoordinates are defined as $Z=(z,\theta^{+},\theta^{-})$
for the holomorphic part and ${\bar Z}=({\bar z},{\bar \theta}^{+},
{\bar \theta}^{-})$ for the anti-holomorphic part, and we denote by
$\phi^{+}$ and $\phi^{-}$ the chiral and anti-chiral superfields which
satisfy the constraints $D_{-} \phi^{+} = {\overline D}_{-}\phi^{+}=0$ and
$D_{+} \phi^{-} = {\overline D}_{+}\phi^{-}=0$, where  the N=2 supercovariant
derivatives are defined as $D_{\pm} = \partial/{\partial{\theta}^{\pm}} +
{1\over 2}\theta^{\mp}{\partial}_z$ and similar expressions for
${\overline D}_{+}$ and ${\overline D}_{-}$.
The chiral and anti-chiral superfields are composed of a complex
boson ${\varphi}^{\pm}$ and a
complex fermion ${\psi}^{\pm}$, ${\bar \psi}^{\pm}$.
In the real basis
${\varphi}^{\pm}={1\over \sqrt2}({\varphi}_1 \pm i{\varphi}_2)$
one of the bosonic parts of the equations of motion, for $\varphi_1$,
in the limit of vanishing fermion fields becomes a sine-Gordon equation and
the other one, for $\varphi_2$, becomes a sinh-Gordon (hyperbolic sine-Gordon)
equation.

  The equations have infinitely many conserved charges at the quantum level as
well as at the classical level.  The quantum theory is constructed by adding a
relevent chiral perturbation to the N=2 free action
$S_{free}$ as
$$\eqalign{
& S = S_{free} - \lambda \int d^2 z \Phi ( z, {\bar z})  \quad , \cr
& \Phi ( z, {\bar z}) = \int  d^2 \theta^{+}( e^{i\beta \phi^{+}}
+ e^{-i\beta \phi^{+}}) + \int d^2 \theta^{-}( e^{i\beta \phi^{-}} +
e^{-i\beta \phi^{-}}) \cr
}\eqno(2)$$
where $\phi^{+}$ and $\phi^{-}$ are free chiral and anti-chiral superfields
satisfying the conditions: \hfill\break
${\overline D}_{+}D_{+}\phi^{+} = 0$ and ${\overline D}_{-}D_{-}\phi^{-} = 0$.
Hence they are decomposed into holomorphic and anti-holomophic
parts $\phi^{\pm}=S^{\pm} + {\bar S}^{\pm}$.

The theory is closely related to the N=2 super conformal field theory.
In the N=2 super CFT, the super energy-momentum tensor with the
Feigin-Fuchs linear term is given as
$$ T = :D_{+}S^{+}D_{-}S^{-}: - i \alpha ({\partial}S^{+} - {\partial}S^{-})
\eqno(3)$$
which contains the N=2 superconformal generators: the U(1) charge,
the supercurrents and the energy-momentum tensor (Virasoro generator).
The central charge is given by $c=3(1-2\alpha^2)$.
In eq.(3) we should put $\alpha = 1/{\beta}$ for the vertex operator
$e^{i\beta S^{\pm}}$ to have the conformal weight ${1\over 2}$ as
the chiral screening operators.
A method to construct the charges from (3) is given in [12,13].
And the first two lower dimensional ones are explicitly shown.
Here we should note that though we applied the CFT technique, our N=2 model
should not be regarded as a perturbed N=2 minimal conformal models, in
contrast to the N=0 and N=1 cases.  This is because, if so, one of the chiral
perturbation term leads to an negative conformal weight.
Rather, the present N=2 sine-Gordon theory is to be interpreted as a super-
renormalizable theory with zero background charge, since $\lambda$ has a mass
dimension [12,13].

  Besides the polynomial charges which assure the integrablity of the theory,
the quantum theory has extra charges of the vertex type.   The vertex type
charges in quantum theories are known in several bosonic and supersymmetric
cases.  In our case, we obtain the extra charges deriving from the following
non-local currents:
$$\eqalign{
J^{\pm}(z) &= \int d \theta^{+} d \theta^{-} :e^{\pm i{1\over \beta}
( S^{+} + S^{-} )}: \cr
&= \bigl ( \pm {1\over {\sqrt{2}\beta}} \partial_z
{\varphi}_2 + {1\over {\beta}^2}\psi^{+} \psi^{-} \bigr )
e^{\pm{i\sqrt{2}\over \beta} \varphi_1} \cr
}\eqno(4)$$
and similarly we have non-local currents for the anti-holomorphic parts
$${\bar J}^{\pm}({\bar z}) = \bigl ( \mp {
1\over {\sqrt{2}\beta}} \partial_{\bar z} {\bar \varphi}_2
+ {1\over {\beta}^2}{\bar \psi}^{+} {\bar \psi}^{-} \bigr )
e^{\mp{i\sqrt{2}\over \beta} {\bar \varphi}_1}
\eqno(5)$$
Under the perturbation term $\lambda\int d^2 w \Phi(w,{\bar w})$ given by (2)
they satisfy the following conservation laws
$$\partial_{\bar z}J^{\pm} = \partial_z H^{\pm} \quad ,
\quad \partial_z {\bar J}^{\pm} = \partial_{\bar z}{\bar H}^{\pm}
\eqno(6)$$
Hence we can construct conserved charges
$$\eqalign{
& Q_{\pm} = \int dz J^{\pm} + \int d{\bar z} H^{\pm} \cr
& {\bar Q}_{\pm} = \int d{\bar z} {\bar J}^{\pm} + \int dz {\bar H}^{\pm} \cr
}\eqno(7)$$
It turns out that these quantum conserved charges generate the $q$-deformation
of $sl(2)$ affine Kac-Moody algebra denoted as $\widehat{sl_{q}(2)}$ which
has been discussed by Bernard and LeClair in the N=0 sine-Gordon case [15].

For the commutation relation between $Q_{+}$ and ${\bar Q}_{-}$, for example,
we
obtain by utilizing the perturbation theory as
$$Q_{+}{\bar Q}_{-} -q^{-2}{\bar Q}_{-}Q_{+}
= \lambda \int_{t} dx \partial_x K
\eqno(8)$$
where the quantum group deformation parameter $q$ is given by
$$ q = -e^{-{i \pi}/\gamma}
\quad , \quad \gamma = \beta^2
\eqno(9)$$
and the $1/\gamma$ is equal to the spin of the conserved charges $Q_{\pm}$.
The quantity $K$ on the right-hand side of (8) is found to be
$$\eqalign{
K = & \bigl (-{1\over \beta^2} \bigr )\bigl \{
\exp  [ i ({{\sqrt{2}}\over {\beta}} - {{\beta}\over {\sqrt{2}}})
\phi_1(x,t) - {{\beta}\over {\sqrt{2}}}\phi_2(x,t)+i\phi_3(x,t)] \cr
& + \exp [ i ({{\sqrt{2}}\over {\beta}} - {{\beta}\over {\sqrt{2}}})
\phi_1(x,t) + {{\beta}\over {\sqrt{2}}}\phi_2(x,t)-i\phi_3(x,t)] \bigr \} \cr
}\eqno(10)$$
where we have bosonized the fermions as $\psi^{\pm}=e^{\pm i \varphi_3}$ and
${\bar \psi}^{\pm}=e^{\pm i {\bar \varphi}_3}$ in order to avoid the subtlety
 of the Grassmann variables. In eq.(10) we denote
$\phi_{i}(x,t)=\varphi_{i}(z) + {\bar \varphi}_{i}({\bar z})$ for i=1,2 and 3.

Defining the toplogical charge,
${\cal T} = {\beta\over 2\pi}{\textstyle \sqrt2} \{\phi_1(x=+\infty) -
\phi_1(x=-\infty)\}$ and taking the soliton configurations where
$\phi_{i}(x=+\infty)=0$ ($i$=1,3) and $\phi_2(x=\pm\infty)=0$, we have
the following boundary condition in accordance with the invariance of the
action [12]:

  i) ${\cal T}$ is an odd integer, $\phi_3(x=-\infty)=\pi$
(mod $2\pi$)  and

ii) ${\cal T}$ is an even integer, $\phi_3(x=-\infty)=0$
(mod $2\pi$). \hfill\break
And thus we find the quantum commutation relation [12]:
$$\eqalign{
&Q_{\pm}{\bar Q}_{\pm} -q^{2}{\bar Q}_{\pm}Q_{\pm} = 0
\cr
&Q_{\pm}{\bar Q}_{\mp} -q^{-2}{\bar Q}_{\mp}Q_{\pm} =a(1-q^{\pm 2{\cal T}})
\cr
[{\cal T}, Q_{\pm} ] &= \pm 2Q_{\pm} \quad , \quad
[{\cal T}, {\bar Q}_{\pm} ] = \pm 2{\bar Q}_{\pm}
\cr
}\eqno(11)$$
where we have defined $a=\lambda/(-\beta^2)$.
{}From this commutation relation we can easily see that the extra charges
generate the quantum affine algebra ${\widehat {sl_{q}(2)}}$ .
Here we should note that it is possible to show by scaling arguments that this
algebra is exact to all orders in perturbation theory.

As we mentioned, since our charges are integrals of the highest
component of their
supermultiplet, they commute with the supercharges. Therefore we have two
commuting symmetries, i.e. ${\widehat {sl_{q}(2)}}$ and N=2 supersymmetry.
This means that the S-matrix is written as a product of \lq\lq minimal
supersymmetric S-matrix \rq\rq and the ordinary sine-Gordon S-matrix
as in the N=1 case discussed by Ahn et al.[11] and by Schoutens [17].
We expect the \lq\lq minimal supersymmetric S-matrix \rq\rq is that of
$$SU(2)_{2} \times SU(2)_{2} / SU(2)_{4} \eqno(12)$$
which is discussed in [11].

The fundamental particles of our model can be represented as the product
of RSOS solitons of the coset model and kinks of the sine-Gordon part.  We
can show that they acctually form the supermultiplets, though we need
suitable truncation to obtain the minimal number of particles.
For this purpose, we consider super-topological charges of N=2 supersymmetry
which are different from the topological charge discussed above.  We will see
the existence of this super-toplogical charges plays a crucial role to reduce
the number of states and to get the minimal number of N=2 states.

The supercharges are holomorphic
without the perturbation, and we should consider the effect of the
perturbation terms.  With the same methods to the bosonic topological charge,
it is easily checked that the superalgebra has the topological modification
[18] as we will see below.

The holomorphic and anti-holomorphic supercurrents are written in terms of
component fields;
$$\eqalign{
& G^{\pm} (z)=\psi^{\pm} (z) \partial_z \varphi^{\pm} (z) \cr
& {\bar G}^{\pm} ({\bar z})={\bar \psi}^{\mp} ({\bar z}) \partial_{\bar z}
{\bar \varphi}^{\mp} ({\bar z}) \cr
}\eqno(13)$$
which obey the conservation laws in perturbation theory as in the extra
non-local currents:
$$\partial_{\bar z}G^{\pm} = \partial_z F^{\pm} \quad ,
\quad \partial_z {\bar G}^{\pm} = \partial_{\bar z}{\bar F}^{\pm}
\eqno(14)$$
where
$$\eqalign{
&F^{\pm} = \lambda \sum_{ab}f^{\pm(ab)}(z){\bar \Phi}^{(ab)}({\bar z}) \cr
&{\bar F}^{\pm} = \lambda \sum_{ab} \Phi^{(ab)}(z){\bar f}^{\pm(ab)}({\bar z})
\cr
}\eqno(15)$$
Here we have written the perturbation term
$\lambda\int d^2 w \Phi(w,{\bar w})$ as
$$\Phi(w,{\bar w})=\sum_{a,b=+,-}\Phi^{(ab)}(w){\bar \Phi}^{(ab)}({\bar w})
\eqno(16)$$
with
$$\eqalign{
&\Phi^{(+ \pm)}(w)= \pm i\beta\psi^{+}(w)
e^{\pm i\beta\varphi^{-}}(w) \cr
&\Phi^{(- \pm)}(w)= \pm i\beta\psi^{-}(w)
e^{\pm i\beta\varphi^{+}}(w) \cr
&{\bar \Phi}^{(+ \pm)}({\bar w})= \pm i\beta{\bar \psi}^{+}({\bar w})
e^{\pm i\beta{\bar \varphi}^{-}}({\bar w}) \cr
&{\bar \Phi}^{(- \pm)}({\bar w})= \pm i\beta{\bar \psi}^{-}({\bar w})
e^{\pm i\beta{\bar \varphi}^{+}}({\bar w}) \cr
}\eqno(17)$$
and $f^{\pm(ab)}(z)$'s and ${\bar f}^{\pm(ab)}({\bar z})$'s are defined
through the operator product expansion:
$$\eqalign{
& G^{\pm}(z)\Phi^{(ab)}(w) \sim {1\over {z-w}}\partial_w f^{\pm(ab)}(w) \cr
& {\bar G}^{\pm}({\bar z}){\bar \Phi}^{(ab)}({\bar w}) \sim
{1\over {{\bar z}-{\bar w}}}\partial_{\bar w} {\bar f}^{\pm(ab)}({\bar w}) \cr
}\eqno(18)$$
Now we define the N=2 supercharges as follows
$${\cal Q}^{\pm}=\int dz G^{\pm} + \int d{\bar z} F^{\pm} \quad , \quad
{\bar {\cal Q}}^{\pm}=\int d{\bar z} {\bar G}^{\pm}
+ \int dz {\bar F}^{\pm}
\eqno(19)$$
and non-vanishing  $f^{\pm(ab)}$'s and ${\bar f}^{\pm(ab)}$'s are given by
$$\eqalign{
& f^{+(-\pm)}=-e^{\pm i\beta\varphi^{+}} \quad , \quad
f^{-(+\pm)}=-e^{\pm i\beta\varphi^{-}} \cr
& {\bar f}^{+(+\pm)}=-e^{\pm i\beta{\bar \varphi}^{-}} \quad , \quad
{\bar f}^{-(-\pm)}=-e^{\pm i\beta{\bar \varphi}^{+}} \cr
}\eqno(20)$$
Perturbation theory leads to the following anti-commutation relation
$$
\bigl \{ {\cal Q}^{\pm}, {\bar {\cal Q}}^{\mp} \bigr \}
= \lambda \int dx \partial_x K^{\pm\mp}\eqno(21)$$
where
$$\eqalign{
K^{\pm\mp} & = \sum_{ab}f^{\pm(ab)}(z){\bar f}^{\mp(ab)}({\bar z})
= e^{i\beta(\varphi^{\pm} + {\bar \varphi}^{\pm})}
+ e^{-i\beta(\varphi^{\pm} + {\bar \varphi}^{\pm})} \cr
& = \bigl [ e^{i{\beta\over \sqrt{2}}(\phi_1(x,t) \pm i\phi_2
(x,t) )} +
e^{-i{\beta\over \sqrt{2}}(\phi_1(x,t) \pm i\phi_2
(x,t) )} \bigr ] \cr
}\eqno(22)$$
while we get
$$\bigl \{ {\cal Q}^{\pm}, {\bar {\cal Q}}^{\pm} \bigr \}=0 \eqno(23)$$
With the same soliton configuration as discussed before we obtain
$$\bigl \{ {\cal Q}^{\pm}, {\bar {\cal Q}}^{\mp} \bigr \}
= \lambda \bigl [ 2 - ( e^{i{\beta\over \sqrt2}\phi_1(-\infty)}
+ e^{-i{\beta\over \sqrt2}\phi_1(-\infty)}) \bigr ] = {\cal T}' \eqno(24)$$
Hence we note that the super topological charge ${\cal T}'$ is determined
by the topological charge ${\cal T}$ as
$$ {\cal T}' = 2\lambda ( 1 - (-1)^{\cal T})\eqno(25)$$
Therefore, when ${\cal T}$ is an odd integer we get ${\cal T}'=4\lambda$,
 on the other hand, if ${\cal T}$ is an even integer we have ${\cal T}'=0$.
Since the soliton and the anti-soliton possess the topological charge
${\cal T}= \pm 1$, they must have non-zero super topological charge
${\cal T}'=4\lambda$.  We will see the physical consequence of this fact in
the later disussion.
In terms of the real basis $Q_1$ and $Q_2$ satisfying
$$\{Q_{1,2}, \bar{Q}_{1,2}\}=2T_{1,2} \eqno(26)$$
${\cal Q}^{\pm}$ and ${\bar {\cal Q}}^{\pm}$ are given by
$${\cal Q}^{\pm}= \textstyle{1\over \sqrt2}(Q_1 \pm i Q_2) \quad , \quad
 {\bar {\cal Q}}^{\pm}= \textstyle{1\over \sqrt2}({\bar Q}_1 \mp i {\bar Q}_2)
\eqno(27)$$
and therefore we have ${\cal T}'=T_1 - T_2 =2T_1 = -2T_2$.

  In order to make our methods clear, we first summarize some well known
results
briefly [19].  The simplest of supersymmetric models is tri-critical Ising
model
which is described by $\phi^{6}$ potential.  The suitably perturbed model
which is still solvable and has supersymmetry is represented as the deformed
$\phi^{6}$ potential which has three minima.  Let us denote these minima by
numbers, -1, 0, 1.  Then fundamental particles, i.e., solitons, correspond
to the classical configurations which go from one minimum to another.
Therefore we have four solitons which we denote $K_{1 0}$, $K_{-1 0}$,
$K_{0 1}$ and $K_{0 -1}$.  Any multi-particle state should have the form as
$$K_{a b}(\theta_{1})K_{b c}(\theta_{2})K_{c d}(\theta_{3}) \cdots \eqno(28)$$
where $\theta$ represents the rapidity of the each particle.

Now we parametrize the S-matrices as
$$K_{a b}(\theta_{1})K_{b c}(\theta_{2})=
\sum_d  S^{bc}_{ad} (\theta_{12}) K_{a d}(\theta_{2}) K_{d c}(\theta_{1})
 \eqno(29)$$
where $a,b,c,d = 0, \pm1$, and $\theta_{12}=\theta_{1}-\theta_{2}$ is the
rapidity difference.

  The supersymmetry is realized on the multi-particle states non-locally.
We denote supercharges as $Q$ and $\bar{Q}$ here.  Then, for N particle
state,
$$\eqalign{
& Q \bigl ( K_{a_{1} a_{2}}(\theta_{1}) K_{a_{2} a_{3}}(\theta_{2}) \cdots
  K_{a_{N-1} a_{N}} \bigr ) = \cr
& \sum_{i=1}^{N}  \beta(a_{i},a_{i+1}) e^{\theta_i/2}
K_{-a_1 -a_2}\cdots K_{-a_{i-1} -a_{i}}
K_{-a_{i} a_{i+1}} K_{a_{i+1} a_{i+2}} \cdots K_{a_{N-1} a_{N}} }
\eqno(30)$$
and the similar expression for $\bar{Q}$ with $\bar{\beta}$.
The functions $\beta$ and $\bar{\beta}$ should be defined so as to realize
superalgebra:
$$\{Q,Q\}=e^{\theta}=P, \quad \{\bar{Q}, \bar{Q}\}=e^{-\theta}=\bar{P},
\quad \{Q,\bar{Q}\}=2T \eqno(31)$$
where we have taken the mass to unity and T is the super-topological charge.
Then
we can choose $$\beta(a,b)=i(a+ib), \quad \bar{\beta}(a,b)=-i(a-ib) \eqno(32)$$
Assuming the commutativity
of the supercharges and S-matrix, Zamolodchikov derived the S-matrix [19].

  The S-matrix can be parametrized as
$$\eqalign{
& K_{0a}(\theta_{1})K_{a0}(\theta_{2})=A_{0}(\theta_{12})K_{0a}(\theta_{2})
K_{a0}(\theta_{1})+A_{1}(\theta_{12})K_{0-a}(\theta_{2})K_{-a0}(\theta_{1})
 \cr
& K_{a0}(\theta_{1})K_{0a}(\theta_{2})=B_{0}(\theta_{12})
K_{a0}(\theta_{2})K_{0a}(\theta_{1}) \cr
& K_{a0}(\theta_{1})K_{0-a}(\theta_{2})=B_{1}(\theta_{12})
K_{a0}(\theta_{2})K_{0-a}(\theta_{1}) \cr
} \eqno(33)$$
where
$$\eqalign{
&A_{0}(\theta)=e^{C\theta}\cosh(\frac{\theta}{4})S(\theta),\quad
A_{1}(\theta)=-ie^{C\theta}\sinh(\frac{\theta}{4})S(\theta) \cr
&B_{0}(\theta)=\sqrt{2}e^{-C\theta}\cosh(\frac{\theta-i\pi}{4})S(\theta)
, \quad
B_{1}(\theta)=\sqrt{2}e^{-C\theta}\cosh(\frac{\theta+i\pi}{4})S(\theta) \cr
} \eqno(34)$$
with
$$\eqalign{
&C=\frac{1}{2\pi i}\log2 \cr
&S(\theta)={1\over \sqrt{\pi}}\prod_{k=1}^{\infty}
\frac{\Gamma(k-\theta/2\pi i)\Gamma(-1/2+k+\theta/2\pi i)}
{\Gamma(1/2+k-\theta/2\pi i)\Gamma(k+\theta/2\pi i)} \cr
}\eqno(35)$$

  Now we proceed to the simplest N=2 super theory, that is, the perturbed
c=1 model which is represented as the coset model (12).
Since this model has two supersymmetries, the S-matrix is represented as
a product of two tri-critical models.  The fundamental particles are
written in the form:
$$K_{a b, c d} \sim K_{a b} K_{c d} \eqno(36)$$

The realization of supersymmetry follows from the realization on the tri-
critical Ising solitons with some modification.  We assume that the first
supercharge operates on the first factor of the right hand side of the
equation just as tri-critical Ising model.  But when the second supercharge
operates on the second factor, we require that the sign of the components of
the first factor should always change.  For illustration, we show how
the supercharge acts on the one-particle state:
$$\eqalign{
& Q_{1}K_{ab, cd}=i(a+ib)e^{\theta/2}K_{-ab, cd}, \quad
  \bar{Q_{1}}K_{ab, cd}=-i(a-ib)e^{-\theta/2}K_{-ab, cd}, \cr
& Q_{2}K_{ab, cd}=i(c+id)e^{\theta/2}K_{-a -b, -cd}, \quad
  \bar{Q}_{2}K_{ab, cd}=-i(c-id)e^{-\theta/2}K_{-a -b, -cd} \cr
}\eqno(37)$$
Therefore the supercharges satisfy
$$\eqalign{
& Q_{1}^{2}=Q_{2}^{2}=e^{\theta}=P,\quad \bar{Q}_{1}^{2}=\bar{Q}_{2}^{2}
=e^{-\theta}=\bar{P},
 \cr
& \{Q_{1,2}, \bar{Q}_{1,2}\}=2T_{1,2}, \quad T_1=-(a^2 - b^2), \quad
T_2=-(c^2 - d^2) \cr
} \eqno(38)$$
and they, otherwise, anticommute with each other.  The reason why we
demand that the sign in the first factor change when we operate the second
charge is to assure the anticommutation relation of $Q_{1}$ and $Q_{2}$.
And we think that this is purely notational matter.  And further, we can show
that the operation of the supercharges commute with the S-matrix.

 We have sixteen fundamental particles.  In order to have apparent
 representation of supersymmetry we need to arrange them to form
supermultiplets.   The tri-critical
Ising solitons can be arranged to form the supermultiplet, i.e. using
$$\eqalign{
B=\frac{1}{\sqrt{2}}\{K_{-10}+K_{10}\}, \quad
F=\frac{1}{\sqrt{2}}\{K_{-10}-K_{10}\} \cr
\bar{B}=\frac{1}{\sqrt{2}}\{K_{0-1}+K_{01}\}, \quad
\bar{F}=\frac{1}{\sqrt{2}}\{K_{0-1}-K_{01}\} }\eqno(39)$$
Fundamental particles that form supermultiplets are
$$ BB, \quad BF, \quad FB, \quad FF, \quad B\bar{B},
\quad B\bar{F}, \quad F\bar{B}, \quad F\bar{F} \eqno(40)$$
and their conjugates, where $BB$, for instance, represents
$$ BB \sim \frac{1}{2}\{K_{-10}+K_{10}\}\{K_{-10}+K_{10}\} $$
following the way of eq.(36)

Since the model has two commuting symmetries, N=2 supersymmetry and
$\widehat{sl_{q}(2)}$ symmetry.  The S-matrix should be factorized into
two parts as in the case of N=1 super sine-Gordon theory [11,20]. It means
$$S_{N2SG} \sim S_{MN2} \times S_{SG} \eqno(41)$$
where $S_{N2SG}$, $S_{MN2}$ and $S_{SG}$ represent the S-matrices of N=2
supersymmetric sine-Gordon, of the minimal N=2 supersymmetric model,
and of sine-Gordon, respectively.

  Now we observe that $S_{MN2}$ is that of the model (12).  Therefore
the fundamental particles are represented as
$$K^{\pm}_{XY} \sim A^{\pm} XY \eqno(42)$$
where $XY$ denotes one of (40) or its conjugates.  The reason for this
observation are; 1) because of the two commuting algebras, the S-matrix
should be factorized, 2) and $S_{MN2}$ should be the ``minimal" N=2
supersymmetric model up to CDD ambiguity[21]. The statement 1) and 2) hold
in other supersymmetric models.

  One question might be whether we have supermulitplets or not.  This is
an old question about (bosonic) sine-Gordon model with a specific value of
the coupling constant, $\beta={2\over \sqrt{3}}$,
which corresponds to c=1 N=2 super CFT.  In the bosonic case, we have only
two fundamental particles which apparently does not form a supermultiplet.
On the other hand, the model is known to correspond to (12) and the
cojectured [11] S-matrix for the model has the supermultiplets of the
fundamental particles. In fact, the particles of sine-Gordon model are
derived by some truncation of the coset model.

  Here we assume we have supermultiplets, because the starting lagrangian
is explicitly supersymmetric.  We might be able to derive restricted model
from our model by truncation, though we do not discuss it here.

Another question may be about CDD ambiguity, though it is related to the
previous question.  At the moment, we do not have any good criterion to
eliminate the ambiguity, except comparing with other methods.  This was
done in bosonic case, but not yet in N=1 and N=2 supersymmetric cases,
as far as we know. This will be discussed elsewhere.

 Now the bosonic sine-Gordon's S-matrix is well known [21].  Let us denote
sine-Gordon
solitons (kink and anti-kink) $A^{+}$ and $A^{-}$.  The S-matrix is,
$${S_{SG}}_{ab;cd}=\frac{U(\theta)}{i \pi} S_{ab;cd} \eqno(43)$$
where $a,b,c,d=\pm$ and ${S_{SG}}_{ab;cd}$ is the scattering amplitude
for the process: $A^{a}+A^{b} \rightarrow A^{c}+A^{d}$ and
$U(\theta)$ given as
$$\eqalign {
& U(\theta)=\Gamma(\frac{1}{\gamma})
\Gamma(1+i\frac{\theta}{\gamma})\Gamma(1-\frac{1}{\gamma} -
i\frac{\theta}{\gamma}) \prod_{n=1}^{\infty} \frac{R_{n}(\theta)R_{n}
(i \pi -\theta)}{R_{n}(0)R_{n}(i \pi)} \cr
&  R_{n}(\theta)=\frac{\Gamma(\frac{2n}{\gamma} +\frac{i \theta}{\gamma})
  \Gamma(1+\frac{2n}{\gamma} +\frac{i \theta}{\gamma})}
{\Gamma(\frac{2n+1}{\gamma}+ \frac{i \theta}{\gamma})
\Gamma(1+\frac{2n-1}{\gamma}+ \frac{i \theta}{\gamma})}\cr
}\eqno(44)$$
with $\gamma$ being given by eq.(9).  And non-zero $S_{ab;cd}$'s are
$$\eqalign{
&S_{++;++}=S_{--;--}=\sinh [\frac{1}{\gamma}(i \pi-\theta)] \cr
&S_{+-;+-}=S_{-+;-+}=\sinh \frac{\theta}{\gamma} \quad , \quad
S_{+-;-+}=S_{-+;+-}=i \sin \frac{\pi}{\gamma} \cr
}\eqno(45)$$
We note that the $S_{ab;cd}$ is the well known solution of Yang-Baxter
equation for the six-vertex model.

  Now we may have ``complete" expression of the S-matrix.  But some
consideration about topological charges leads to reduction of the number of
fundamental particles.  Because of the form of the potential without fermions,
we expect kink and anti-kink solutions as in the case of bosonic and N=1
supersymmetric cases.  And the topological charges are discussed in [12,13]
and in the beginning of this paper.  Since kink and anti-kink solutions have
odd topological charge, their super topological charges should be non-zero.
This restricts $XY$ in eq.(40) to one of
$$B{\bar B}, \quad B{\bar F}, \quad F{\bar B}, \quad F{\bar F} \eqno(46)$$
which possess $T_1=-T_2=-1$ and their conjugates with $T_1=-T_2=1$.
For remaining states the super-topological charges vanish.
In fact, particles of (46) and their conjugates form a closed set, i.e.
we can restrict on-shell states to the set and other states cannot appear
in the final state.  Of course, this is another expression of the conservation
laws of super topological charges.   Therefore we have four supermultiplets;
those of
$$A^{\pm}B{\bar B},\quad A^{\pm}\bar{B}B \eqno(47)$$.

Now we present explict results for minimal N=2 supersymmetric parts of
S-matrices.  First we write down the in-states of tricritical Ising soliton
and anti-soliton scattering in terms of the out-states. This can be
provided by the matrices ${\cal A}$ and ${\cal B}$ given below. And then,
by taking the tensor product of these in-states
we can construct the S-matrices for soliton-antisoliton scattering for
N=2 supersymmety as follows
$$\eqalign{
|X{\bar Y}(\theta_1){\bar V}W(\theta_2)\rangle_{in} &=
|X(\theta_1){\bar V}(\theta_2)\rangle_{in}
|{\bar Y}(\theta_1)W(\theta_2)\rangle_{in} \cr
& = \sum_{X',{\bar Y}',{\bar V}', W'}
{\cal B}_{X{\bar V},X'{\bar V}'}(\theta_{12})
{\cal A}_{{\bar Y}W,{\bar Y}'W'}(\theta_{12})
|X'{\bar Y}'(\theta_2){\bar V}'W'(\theta_1)\rangle_{out} \cr
}\eqno(48)$$
where $X$, $Y$, $V$ and $W$ run over $B$ and $F$. In the above equation the
tricritical Ising soliton amplitudes ${\cal A}$ and ${\cal B}$ are given by
$$
{\cal A} =
\left(\begin{array}{cccc}
A_{+} & A_{+} & 0 & 0 \\
A_{+} & A_{+} & 0 & 0 \\
0 & 0 & A_{-} & A_{-}  \\
0 & 0 & A_{-} & A_{-}
\end{array}\right)
\hspace{1 cm}
{\cal B} =
\left(\begin{array}{cccc}
B_{+} & B_{-} & 0 & 0 \\
B_{-} & B_{+} & 0 & 0 \\
0 & 0 & B_{+} & B_{-}  \\
0 & 0 & B_{-} & B_{+}
\end{array}\right)
\eqno(49)$$
where the row and the column of ${\cal A}$ are arranged in the order:
${\bar B}B$, ${\bar F}F$, ${\bar B}F$ and${\bar F}B$
, while those of ${\cal B}$ are in the order: $B{\bar B}$, $F{\bar F}$,
$B{\bar F}$ and $F{\bar B}$ . In (49)
$A_{\pm}(\theta)$ and $B_{\pm}(\theta)$ are obtained from (34) as
$$\eqalign{
& A_{+}(\theta) = \textstyle{1\over 2}(A_{0}(\theta) + A_{1}(\theta))
= \textstyle{1\over 2}e^{C\theta}(\cosh{{\theta\over 4}}
-i \sinh{{\theta\over 4}})S(\theta) \cr
& A_{-}(\theta) = \textstyle{1\over 2}(A_{0}(\theta) - A_{1}(\theta))
= \textstyle{1\over 2}e^{C\theta}(\cosh{{\theta\over 4}}
+i \sinh{{\theta\over 4}})S(\theta) \cr
& B_{+}(\theta) = \textstyle{1\over 2}(B_{0}(\theta) + B_{1}(\theta))
= e^{-C\theta}\cosh{{\theta\over 4}}S(\theta) \cr
& B_{-}(\theta) = \textstyle{1\over 2}(B_{0}(\theta) - B_{1}(\theta))
= - i e^{-C\theta}\sinh{{\theta\over 4}}S(\theta) \cr
}\eqno(50)$$
For illustration, we consider the scattering of $B{\bar B}$
and ${\bar B}B$ as follows:
$$\eqalign{
&|B{\bar B}(\theta_1){\bar B}B(\theta_2)\rangle_{in} =
|B(\theta_1){\bar B}(\theta_2)\rangle_{in}
|{\bar B}(\theta_1)B(\theta_2)\rangle_{in} \cr
& = B_{+}A_{+}|(B{\bar B})({\bar B}B)\rangle_{out}
+ B_{-}A_{+}|(F{\bar B})({\bar F}B)\rangle_{out} \cr
& + B_{+}A_{+}|(B{\bar F})({\bar B}F)\rangle_{out}
+ B_{-}A_{+}|(F{\bar F})({\bar F}F)\rangle_{out} \cr
}\eqno(51)$$
Hence, for example, we obtain the following result:
$$\eqalign{
& S(B{\bar B} + {\bar B}B \rightarrow B{\bar B} + {\bar B}B) \cr
& = S(B{\bar B} + {\bar B}B \rightarrow B{\bar F} + {\bar B}F) \cr
& =B_{+}A_{+}
=\textstyle{1\over 4}(1+\cosh{\theta\over 2}- i\sinh{\theta\over 2})
S^2(\theta) \cr
}\eqno(52)$$
The S-matrices for the processes: ${\bar X}Y +V{\bar W} \rightarrow
{\bar X}'Y' +V'{\bar W}'$ are obtained by crossing.
The total S-matrices are given as a product of above N=2 supersymmetry parts
wih the ordinary sine-Gordon S-matrix factors as follows:
$$S_{N2SG}(\theta)=S_{MN2}(\theta) \times S_{SG}(x=e^{\theta/\gamma},
q=-e^{-i\pi/\gamma})\eqno(53)$$

Here we have a comment on the correspondence between the coset and the bosonic
sine-Gordon theories.
As we mentioned before, the N=2 supersymmetry is realized for the special
value of the coupling constant, $\beta={2\over \sqrt{3}}$.
By setting $\gamma=2$ corresponding to the above value of $\beta$ in
eqs.(43)-(45)(or setting $\gamma=16\pi$ in the S-matrix formula of ref.[21]
where the normalization of $\gamma$ differs from ours by a factor $8\pi$)
for ordinary bosonic sine-Gordon theory, we obtain the
scattering matrices for soliton $A$ and anti-soliton ${\bar A}$
$$\eqalign{
&A(\theta_1){\bar A}(\theta_2)=S_{T}(\theta_{12}){\bar A}(\theta_2)A(\theta_1)
+S_{R}(\theta_{12})A(\theta_2){\bar A}(\theta_1) \cr
& A(\theta_1)A(\theta_2)=S_{0}(\theta_{12})A(\theta_2)A(\theta_1) \quad ,
\quad
{\bar A}(\theta_1){\bar A}(\theta_2)=S_{0}(\theta_{12})
{\bar A}(\theta_2){\bar A}(\theta_1) \cr
}\eqno(54)$$
where $S_T$ and $S_R$ are transmission and reflection amplitudes for soliton-
antisoliton scattering and $S_0$ denotes the amplitude for identical soliton
scattering
$$\eqalign{
& S_{T}(\theta)= -i\sinh{\theta\over 2}S^2(\theta) \cr
& S_{0}(\theta)= \cosh{\theta\over 2}S^2(\theta) \cr
& S_{R}(\theta)= S^2(\theta) \cr
}\eqno(55)$$
What is remakable here is that we can recover these results by expressing
the soliton states in terms of the coset soliton states as
$$\eqalign{
A &= {\bar B}B -{\bar F}B + {\bar B}F + {\bar F}F
+B{\bar B} + F{\bar B} -B{\bar F} + F{\bar F} \cr
& = K_{01}K_{-10} + K_{01}K_{10} + K_{-10}K_{0-1} + K_{-10}K_{01}
+ K_{0-1}K_{-10} - K_{0-1}K_{10} \cr
& - K_{10}K_{0-1} + K_{10}K_{01} \cr
}\eqno(56)$$
This strongly supports the validity of our calculation for S-matrices of
N=2 SG soliton scattering.

Now some remarks on N=2 SG breathers are in order.
 As in the case of kinks, breathers can be composed
of those of two tri-critical Ising models and bosonic sine-Gordon.  At first
we need the supermultiplets of tricritical Ising model.
In our case, the breathers are represented as
$$B_{n}x_{n}y_{n} \eqno(57)$$
where $B_{n}$ is a breather of sine-Gordon and $x_{n}$ and $y_{n}$ represent
$\phi_{n}$ or $\psi_{n}$ which are breathers of the tricritical Ising model
[20].
The supersymmetry is realized on the particles
with no topological charge. We should easily construct soliton-breather or
breather-breather S-matrices in the same manner described above.

We have considered S-matrix of N=2 supersymmetric sine-Gordon theory.
As investigated explicitly, the theory has two quantum symmetries,
$\widehat{sl_{q}(2)}$ and N=2 supersymmetry.  Utilizing the symmetry we
were able to costruct the S-matrix explicitly.  One interesting point is the
topological charges of the two symmetries.  We derived the relation of three
charges (one for $\widehat{sl_{q}(2)}$ and two for N=2 SUSY), which is
originally discussed by Witten and Olive [18]. It turns out that the
relation between the topological and super-toplogical charges
is essential when we restrict the fundamental particles from the abstract
realization theory.

It is interesting to investigate this model in terms of other methods, like
WKB methods to compare with the results.  We point out the fact that we have
some difficulty in the investigation of supersymmetric solvable models, like
inverse scattering method, because of the exsistence of fermions.  But in N=2
super cases, whose investigation has just started, we can bosonize the
fermions to avoid the problem.  Or in another word, we can start from the
lagrangian which has only bosonic fields and is supersymmetric.  We think this
is an interesting investigation to see how the quatum group method works.
Our model is the simplest model and we believe it can be extended to any N=2
supersymmetric models [22-24]. In this regards, it would be intriguing to
extend the present analysis to the N=2 supersymmetric version of Toda field
theories [25-30].
We have one more comment about ``supersymmetry" ( see refs.[16,17,22,23]).
We think it becomes
clear that S-matrix of N-supersymmetric models can be constructed simply by
making the N product of that of the N=1 supersymmetric model (tri-critical
Ising model) and some bosonic part, up to CDD ambiguity.  We are interested
in whether there is a model for arbitrary N.

\bigskip

We thank T.~Inami and S.~K.~Yang for useful discussions. One of us (K.~K.)
would like to thank the members of high-energy theory group at Purdue
University and the theory group in College of Liberal Arts and Sciences,
Kyoto University for their kind hospitality.
This work was partially supported by the Grant-in-Aid for Scientific Research
from the Ministry of Education, Science and Culture (\#63540216).

\newpage
\vspace{0.8 cm}

\vspace{0.8 cm}
\centerline{\large \bf  References}
\vspace{0.8 cm}

\begin{description}

\item{[1]}
A.~B.~Zamolodchikov, Sov.~J.~Nucl.~Phys.~{\bf 44}(1986)529; Reviews in
\hfill\break
Mathematical Physics {\bf 1}(1990)197.

\item{[2]}
T.~Eguchi and S.~K.~Yang, Phys.~Lett.~{\bf 224B}(1989)373 and {\bf 235B}(1990)
282.

\item{[3]}
F.~A.~Smirnov, Int.~J.~Mod.~Phys.~{\bf A4}(1989)4213; Nucl.~Phys.~{\bf B337}
(1990)156.

\item{[4]}
A.~LeClair, Phys.~Lett.~{\bf 230B}(1989)103.

\item{[5]}
N.~Reshetikhin and F.~Smirnov, Commun.~Math.~Phys.~{\bf 131}(1990)157.

\item{[6]}
V.~A.~Fateev and A.~B.~Zamolodchikov, Int.~J.~Mod.~Phys.~{\bf A6}(1990)1025.

\item{[7]}V.~G.Drinfel'd, Sov.~Math.~Dokl.~{\bf 32}(1985)254.

\item{[8]}M.~Jimbo, ~Lett.~Math.~Phys.~{\bf 10}(1985)63.

\item{[9]}
D.~Bernard and A.~LeClair, Nucl.~Phys.~{\bf B340}(1990)721; Phys.~Lett.
\hfill\break {\bf 247B}(1990)309.

\item{[10]}P. ~Mathieu, Nucl.~Phys.~{\bf B336}(1990)338.

\item{[11]}
C.~Ahn, D.~Bernard and A.~LeClair, Nucl.~Phys.~{\bf B346}(1990)409.

\item{[12]}
K.~Kobayashi and T.~Uematsu, Phys.~Lett.~{\bf 264B}(1991)107.

\item{[13]}
K.~Kobayashi, T.~Uematsu and Y.-Z.~Yu, \lq\lq Quantum Conserved Charges
in N=1 and N=2 Supersymmetric Sine-Gordon Theories \rq\rq KUCP-29/91;
 PURD-TH-91-03.

\item{[14]}
T.~Inami and H.~Kanno, Commun.~Math.~Phys.~{\bf 136}(1991)519; Nucl.~Phys.
\hfill\break {\bf B359}(1991)201.

\item{[15]}
D.~Bernard and A.~LeClair, \lq\lq Quantum Group Symmetries and Non-Local
Currents in 2D QFT \rq\rq, CLNS-90-1027(1990);SPhT-90/144(1990).

\item{[16]}
R.~Shankar and E.~Witten, Phys.~Rev.~{\bf D17}(1978)2134; Nucl.~Phys.
\hfill\break {\bf B141}(1978)525.

\item{[17]}
K.~Schoutens, Nucl.~Phys.~{\bf B344}(1990)665.

\item{[18]}
E.~Witten and D.~Olive, Phys.~Lett.~{\bf 78B}(1978)97.

\item{[19]}
A.~B.~Zamolodchikov, \lq\lq Fractional-Spin Integrals of Motion in Perturbed
CFT \rq\rq , Moscow preprint (1989).

\item{[20]}
C.~Ahn, Nucl.~Phys.~{\bf B354}(1991)57.

\item{[21]}
A.~B.~Zamolodchikov and A.~B.~Zamolodchikov, Ann.~Phys.~(NY)~{\bf 120}
\hfill\break (1979)253.

\item{[22]}
P.~Fendley, S.~D.~Mathur, C.~Vafa and N.~P.~Warner,
Phys.~Lett.~{\bf 243B}(1990)257.

\item{[23]}
P.~Fendley, W.~Lerche, S.~D.~Mathur and N.~P.~Warner, Nucl.~Phys.~{\bf B348}
(1991)66.

\item{[24]}
P.~Mathieu and M.~A.~Walton, \lq\lq Integrable Perturbations of N=2
Superconformal Minimal Models \rq\rq , Quebec preprint(1990).

\item{[25]}
T.~J.~Hollowood and P.~Mansfield, Phys.~Lett.~{\bf 226B}(1989)73.

\item{[26]}
J.~Evans and T.~Hollowood, Nucl.~Phys.~{\bf B352}(1991)723.

\item{[27]}
H.~W.~Braden, E.~Corrigan, P.~E.~Dorey and R.~Sasaki, Nucl.~Phys.~{\bf B338}
(1990)689.

\item{[28]}
T.~Nakatsu, Nucl.~Phys.~{\bf B356}(1991)499.

\item{[29]}
G.~W.~Delius, M.~T.~Grisaru, S.~Penati and D.~Zanon, Nucl.~Phys.~{\bf B359}
(1991)125.

\item{[30]}
S.~Komata, K.~Mohri and H.~Nohara, Nucl.~Phys.~{\bf B359}(1991)168.

\end{description}

\end{document}